\documentclass[aps,prl,floats,nofootinbib]{revtex4}
\usepackage{psfig}

\draft

\def\lsim{\mathrel{\raise.3ex\hbox{$<$\kern-.75em\lower1ex\hbox{$\sim$}}}}
\def\gsim{\mathrel{\raise.3ex\hbox{$>$\kern-.75em\lower1ex\hbox{$\sim$}}}}

\begin{document}

\hbox to \textwidth{\hfill\vbox{
\hbox{MADPH--04--1367}
\hbox{ILL-(TH)-04-3}
\hbox{NSF-KITP-04-38}
\hbox{hep-ph/0404182}}}

\title { Scale of Quantum Gravity }
\author{Tao Han$^{1,3,4}$ and Scott Willenbrock$^{2,3}$ }
\affiliation{
$^{1}$Department of Physics, University of Wisconsin
1150 University Avenue, Madison, WI\ \ 53706, USA \\
\smallskip
$^{2}$Department of Physics, University of Illinois at Urbana-Champaign
1110 West Green Street, Urbana, IL\ \ 61801, USA \\
\smallskip
$^{3}$Kavli Institute for Theoretical Physics, University of California
Santa Barbara, CA\ \ 93106, USA \\
\smallskip
$^{4}$Institute of Theoretical Physics, Academia Sinica,
Beijing 100080, P.~R.~China }


\begin{abstract}
In the effective field theory of quantum gravitation coupled to
$N_s$ scalars, $N_f$ fermions, and $N_V$ vectors, tree unitarity
is violated at an energy squared of $E_{CM}^2 = 20(G_NN)^{-1}$,
where $N\equiv {2\over 3}N_s+N_f+4N_V$ and $G_N$ is Newton's
constant. This is related to radiative corrections proportional to
$G_NNE^2$ (where $E$ is the typical energy), due to loops of such
particles.  New physics must enter before $E_{CM}\approx 6 \times
10^{18}$ GeV in the standard model, and $4 \times 10^{18}$ GeV in
the minimal supersymmetric standard model.
\end{abstract}

\maketitle

\smallskip

Shortly after discovering the fundamental constant of quantum mechanics that
now bears his name, Planck noticed that it may be combined with the
fundamental constants of relativity and gravitation to make a constant with the
units of mass.  This mass also bears his name, $M_{Pl}\equiv (\hbar
c/G_N)^{1/2}\approx 10^{19}$ GeV/c$^2$.  While he could not know the physical
significance of this extremely large mass, he did remark that it could be
considered as the fundamental unit of mass.

A century later, we still do not know the true physical
significance of the Planck mass.  Instead, we regard the Planck
mass in the context of an effective quantum field theory of
gravitation \cite{Weinberg:1978kz,Donoghue:dn,Burgess:2003jk}.  At
leading order, this theory is simply Einstein's theory of general
relativity. Beyond leading order, there are corrections to the
predictions of Einstein's theory proportional to powers of
$E^2/M_{Pl}^2$, where $E$ is the typical energy of the process
under consideration (here and henceforth we work with units where
$\hbar=c=1$, for convenience). Thus the expansion parameter of
this effective theory is proportional to $G_NE^2$.
This expansion, and hence the
usefulness of the effective theory, breaks down at energies of
order the Planck mass.

In this paper we endeavor to make this statement more precise.  We
use tree unitarity to determine the energy at which the effective
theory of quantum gravity breaks down. While dimensional analysis
dictates that the theory breaks down at energies of order
$M_{Pl}$, it does not reveal the presence of dimensionless
constants. Does the expansion break down at an energy close to the
Planck mass, $M_{Pl}\approx 10^{19}$ GeV, the reduced Planck mass,
$M_{Pl}/\sqrt{8\pi} \approx 2.4\times 10^{18}$ GeV (which appears
naturally in Einstein's theory \cite{Polchinski:1998rr}), or
perhaps some altogether different energy?

When the effective theory breaks down, a new description of the
physics becomes appropriate.  This new description may become
relevant before the energy at which the effective theory breaks
down.  Thus the energy at which the effective theory breaks down
places an upper bound on the energy at which new physics must
enter.  Unitarity has been used to determine the energy at which
other effective field theories break down, such as the Fermi
theory of the weak interaction \cite{Lee:qv}, the
spontaneously-broken electroweak theory (with no Higgs boson)
\cite{Chanowitz:1985hj}, and the standard model in higher
dimensions \cite{He:1999qv}.

%
\begin{figure}[tb]
\begin{center}
    \mbox{\psfig{file=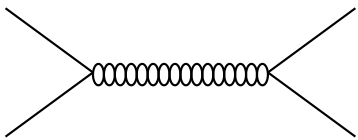,width=5cm} }
\end{center}
    \caption {Scattering of elementary particles via $s$-channel
 graviton exchange.}
    \label{tree}
\end{figure}

The partial-wave amplitude for a $2\to 2$ elastic scattering
amplitude satisfies the unitarity condition $|{\rm Re}\, a_J|\le
1/2$, where $J$ is the total angular momentum. If the
leading-order contribution to the amplitude violates this
condition, then the expansion cannot converge and the effective
theory is useless.  Our approach is thus to calculate all $2\to 2$
scattering amplitudes in quantum gravity and to ask at what energy
they violate unitarity.

The standard model contains scalar, fermion, and vector fields,
with couplings to gravity that depend only on the particles'
masses and momenta.  At energies of order the Planck mass, the
particles' masses are negligible. For convenience, we calculate
the scattering amplitudes in the symmetric phase of the
electroweak theory, where all particle masses vanish.

We begin with the scattering of scalar particles, $s\bar s\to
s'\bar s'$. There are two such (complex) scalar particles in the
standard model, corresponding to the two components of the Higgs
doublet field.  We consider  the case $s \neq s'$, in which
there is only an $s$-channel diagram, as shown in Fig.~\ref{tree}.
There are two reasons for doing so. First, amplitudes which include a
$t$- or $u$-channel diagram do not yield a well-defined
partial-wave amplitude, since the integration over the scattering
angle diverges at $\theta = 0$ or $\pi$.  This corresponds to an
infrared singularity due to the exchange of an on-shell graviton,
and is an irrelevant complication which we choose to avoid.
Second, the fraction of processes in which $s=s'$ diminishes as
the number of particles increases. There are many fields in the
standard model, so neglecting this small subset of processes is a
good approximation.  It is for these same reasons that we neglect
processes of the type $ss'\to ss'$.

The scattering of scalar particles is actually the most subtle case, since
their coupling to gravitons is not fixed uniquely by their mass and momenta at
leading order. The coupling of a scalar field to gravity is given at leading
order by
\begin{eqnarray}
\sqrt{-g}\ {\cal L}= \frac{1}{16\pi
G_N}\sqrt{-g}\ (-2\lambda+R) +
 \sqrt{-g}\ (g^{\mu\nu}\partial_\mu\phi^\dagger
\partial_\nu\phi - m^2|\phi|^2 + aR|\phi|^2)+...\ \ \
\label{eq:lagrangian}
\end{eqnarray}
where $g$ is the determinant of the metric, $\lambda$ is the
cosmological constant, $R$ is the Ricci scalar, $m$ is the scalar
mass, and $a$ is a free parameter.  We do not know the value of
$a$ in the standard model, so we leave it as a free parameter. It
is only for a scalar field that such an ambiguity exists.  The
case $m=0$, $a=-1/12$ corresponds to the conformal limit of the
theory \cite{Callan:ze}.

The amplitude for $s$-channel scalar-scalar scattering, at energies much
greater than the scalar mass, is
\begin{equation}
{\cal A}(s\bar s\to s'\bar s')=-{4\over 3}\pi
G_NE_{CM}^2\left[d^2_{0,0}-(1+12a)^2d^0_{0,0}\right]
\end{equation}
where $E_{CM}$ is the total center-of-momentum energy and
$d^J_{\mu,\mu'}$ is the Wigner $d$ function, where $J$ is the
total angular momentum, $\mu=\lambda-\bar\lambda$ is the
difference of the helicities of the initial-state particles, and
$\mu'=\lambda'-\bar\lambda'$ is the difference of the helicities
of the final-state particles. The first term corresponds to $J=2$
exchange, while the second corresponds to $J=0$.  The $J=0$
component of a virtual graviton couples to the trace of the
energy-momentum tensor, which explains why the $J=0$ term vanishes
in the conformal limit ($a=-1/12$), where $T^\mu_\mu=0$.

The presence of a $J=0$ component of a virtual graviton does not
violate any physical principles.  While a real graviton must
contain only helicity $\pm 2$ components
\cite{Veltman:vx,'tHooft:bx,Feynman:kb}, a virtual graviton can
contain a $J=0$ component. There is no violation of
angular-momentum selection rules in quantum gravity, as recently
claimed in Ref.~\cite{Datta:2003kn}.  This claimed was based on
the assumption that a virtual graviton cannot have a $J=0$
component.

\begin{table*}[tbh]
\begin{center}
\begin{tabular}{|c|c|c|c|c|c|}
\hline
 & $s'\bar s'$ & $f'_+\bar f'_- $ & $ f'_-\bar f'_+ $ & $V'_+ V'_-$ & $V'_- V'_+$ \\
\hline $s\bar s$ $\to $ & ${2}/{3}d^2_{0,0} -
{2}/{3}(1+12a)^2d^0_{0,0}$ & $ \sqrt{2/3}\ d^2_{0,1}$
& $\sqrt{2/3}\ d^2_{0,-1}$  & $ 2\sqrt{2/3}\ d^2_{0,2} $  & $ 2\sqrt{2/3}\ d^2_{0,-2} $\\
\hline $f_+\bar f_-\to $ &
                $ \sqrt{2/3}\ d^2_{1,0}$ & $d^2_{1,1} $
& $  d^2_{1,-1} $ &  $2\ d^2_{1,2}$ & $2\ d^2_{1,-2}$ \\
\hline $f_-\bar f_+\to $ &
                $\sqrt{2/3}\ d^2_{-1,0}$ & $ d^2_{-1,1} $
& $ d^2_{-1,-1} $ &   $2\ d^2_{-1,2}$ & $ 2\ d^2_{-1,-2}$ \\
\hline $V_+ V_-\to $ &  $2\sqrt{2/3}\ d^2_{2,0}$ &  $ 2\
d^2_{2,1}$ &$
 2\ d^2_{2,-1}  $ & $4\ d^2_{2,2}$ & $4\ d^2_{2,-2}$ \\
\hline $V_- V_+ \to $ &  $2\sqrt{2/3}\ d^2_{-2,0}$ &  $ 2\
d^2_{-2,1}$ &$
 2\ d^2_{-2,-1}  $ & $4\ d^2_{-2,2}$ & $4\ d^2_{-2,-2}$ \\
\hline \hline
\end{tabular}
\caption{Scattering amplitudes for scalars, fermions, and vector
bosons via $s$-channel graviton exchange in terms of the Wigner
$d$ functions. The subscripts on the particles indicate their
helicities. All particle masses have been neglected. An overall
factor $-2\pi G_N E_{CM}^2$ has been extracted from the
amplitudes.} \label{tab:amplitudes} \end{center}
\end{table*}

We calculate the scattering amplitude, in the high-energy limit,
of all $s$-channel processes with initial- and final-state
scalars, fermions, and vector bosons.  This covers all the
particles of the standard model, as well as the supersymmetric
standard model (excluding gravitons and gravitinos).  Only certain
amplitudes are allowed, due to helicity conservation in the
gravitational interaction of the external particles in the
massless limit. The nonzero amplitudes are given in
Table~\ref{tab:amplitudes}.\footnote{These amplitudes are
calculated using the Feynman rules of Ref.~\cite{Han:1998sg}, the
spinors of Ref.~\cite{Willenbrock:2002ta}, and the polarization
vectors of Ref.~\cite{Hagiwara:1985yu}.  The Wigner $d$ functions
may be found in Ref.~\cite{Hagiwara:fs}.}

We now convert these amplitudes to partial-wave amplitudes, $a_J$,
defined by
\begin{equation}
{\cal A}=16\pi \sum_J (2J+1)a_J d^J_{\mu,\mu'}\;.
\end{equation}
Thus the matrix of $J=2$ partial-wave amplitudes is proportional
to the matrix of Table~\ref{tab:amplitudes}, but with the Wigner
$d$ functions discarded, and the term proportional to $d^0_{0,0}$
eliminated from the $s\bar s\to s'\bar s'$ amplitude.

We wish to diagonalize the matrix of partial-wave amplitudes when
there are $N_s$ scalars, $N_f$ fermions, and $N_V$ vector bosons.
To simplify the task, let us combine the states $f_+\bar f_-$ and
$f_-\bar f_+$ into the single state $f_-\bar f_+$, where $f_-$
denotes any of the fifteen left-handed states of one generation
($u_R,$$u_G,$$u_B,$$\bar u_R,\bar u_G,$$\bar u_B,d_R,d_G,d_B,$$\bar d_R,\bar
d_G,\bar d_B,e^-,e^+,\nu$).\footnote{We are assuming that
neutrinos are Majorana fermions, and that the left-handed state
$\bar\nu$ is not present in the theory.} Since all fermions have
the same gravitational interactions (in the massless limit), we
should consider the normalized state obtained by summing over all
fermion-antifermion states, $(1/\sqrt{N_f})\Sigma f_-\bar f_+$.
The same is true of the scalars and the vector bosons.
Furthermore, the states $V_+V_-$ and $V_-V_+$ are identical in the
partial-wave amplitudes, so we need only consider the former.

We are thus led to construct the matrix of partial-wave amplitudes
in the basis of states $(1/\sqrt{N_s})\Sigma s\bar s$,
$(1/\sqrt{N_f})\Sigma f_-\bar f_+$, and $(1/\sqrt{N_V})\Sigma
V_+V_-$, as shown in Table~\ref{tab:partial}.  Each element in the
matrix corresponds to a sum of many identical amplitudes.  For
example, the element in the first row, second column, sums the
amplitudes for $N_s$ pairs of scalars scattering into $N_f$ pairs
of fermions.  Hence the element is proportional to $N_sN_f$ times
the normalization factors $1/\sqrt{N_s}$ and $1/\sqrt{N_f}$ from
the basis states.\footnote{The diagonal elements of this matrix
include processes in which the initial and final states are
identical.  We are ignoring the $t$- and $u$-channel contributions
to these amplitudes, for reasons discussed above.  The $s$-channel
contribution by itself is gauge invariant.}

\begin{table*}[tbh]
\begin{center}
\begin{tabular}{|c|c|c|c|}
\hline
 & ${1\over {\sqrt N_s}}\Sigma s'\bar s'$ & ${1\over {\sqrt N_f}}\Sigma f'_-\bar f'_+ $
 & $ {1\over {\sqrt N_V}}\Sigma V'_+ V'_-$ \\
\hline ${1\over {\sqrt N_s}}\Sigma s\bar s$ $\to $ & ${2}/{3}N_s$
& $ \sqrt{2/3}\sqrt{N_sN_f}$
& $ 2\sqrt{2/3}\sqrt{N_sN_V}$ \\
\hline ${1\over {\sqrt N_f}}\Sigma f_-\bar f_+\to $ & $\sqrt{2/3}\sqrt{N_sN_f}$ & $ N_f $
& $2\sqrt{N_fN_V}$ \\
\hline ${1\over {\sqrt N_V}}\Sigma V_+ V_-\to $ &  $2\sqrt{2/3}\sqrt{N_sN_V}$ &  $ 2\sqrt{N_fN_V}
$ & $4N_V$  \\
\hline \hline
\end{tabular}
\caption{$J=2$ partial-wave amplitudes for $N_s$ scalars, $N_f$
fermions, and $N_V$ vector bosons via $s$-channel graviton
exchange. The subscripts on the particles indicate their
helicities.  An overall factor $-G_N E_{CM}^2/40$ has been
extracted from the partial-wave amplitudes.} \label{tab:partial}
\end{center}
\end{table*}

Unitarity demands that the eigenvalues of the matrix of partial-wave
amplitudes satisfy
\begin{equation}
{\rm Im}\; a_J \ge |a_J|^2 \label{eq:unitarity}
\end{equation}
which implies that $|{\rm Re}\; a_J|\le 1/2$. The matrix in
Table~\ref{tab:partial} has only one nonzero eigenvalue,
\begin{equation}
a_2^{(1)}= -{1\over 40}G_N E_{CM}^2\left({2\over
3}N_s+N_f+4N_V\right)\;, \label{eq:eigenvalue}
\end{equation}
corresponding to the (unnormalized) eigenvector $\sqrt{2\over
3}\Sigma s\bar s+\Sigma f_-\bar f_++2\Sigma V_+ V_-$. Tree
unitarity is therefore violated at an energy squared of
\begin{equation}
E_{CM}^2=20(G_N N)^{-1}, \;\;\;\;\;N\equiv {2\over 3}N_s+N_f+4N_V\;.
\label{bound}
\end{equation}
In the standard model with one Higgs doublet and three generations
of fermions, $N_s=2$, $N_f=45$, and $N_V=12$.  Thus tree unitarity
is violated at an energy $E_{CM}=\sqrt{60/283}\;G_N^{-1/2}\approx
6\times 10^{18}$ GeV. This is about one half of the Planck mass.

If nature is supersymmetric, then there are many more particles to
consider, since every standard-model particle has a
superpartner.\footnote{The superpartners of the vector bosons are
Majorana fermions, so $\bar f=f$ in the scattering amplitudes. The
same is true of Majorana neutrinos.}  In the minimal
supersymmetric standard model with two Higgs doublets,
$N_s=45+4=49$, $N_V=12$, and $N_f=N_s+N_V=61$. Unitarity is
violated at an energy $E_{CM}=\sqrt{12/85}\;G_N^{-1/2}\approx
4\times 10^{18}$ GeV.  This is about one third of the Planck mass.

Alternatively, one could consider the $J=0$ partial-wave amplitude,
which only receives a contribution from scalars.  One finds that tree
unitarity is violated at an energy squared of
\begin{equation}
E_{CM}^2 = 6 (G_N N_s)^{-1}/(1+12a)^2.
\end{equation}
We do not know the value of $a$, so let us choose $a=0$ for
illustrative purposes. With this choice, tree unitarity is
violated in the standard model at $E_{CM} = \sqrt 3\ G_N^{-1/2}$,
while in the minimal supersymmetric standard model it is violated
at $E_{CM} = \sqrt{6}/7\; G_N^{-1/2}\approx 4\times 10^{18}$ GeV.

\begin{figure}[tb]
\begin{center}
    \mbox{\psfig{file=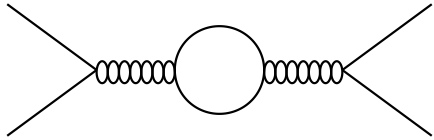,width=6cm} }
\end{center}
    \caption {One-loop correction to the scattering of elementary particles  via $s$-channel
graviton exchange. The loop contains a sum over all elementary particles.}
    \label{loop}
\end{figure}

It is easy to understand the origin of the factor $N\equiv {2\over
3}N_s+N_f+4N_V$ that appears in the partial-wave amplitude,
Eq.~(\ref{eq:eigenvalue}). Consider the $s$-channel elastic
scattering of the state given by the eigenvector $\sqrt{2\over
3}\Sigma s\bar s+\Sigma f_-\bar f_++2\Sigma V_+ V_-$, represented
by the tree diagram in Fig.~\ref{tree}. The one-loop correction to the
amplitude due to particles other than gravitons is given by the
diagram in Fig.~\ref{loop}, where the loop contains scalars, fermions, and
vector bosons. Elastic unitarity, via Eq.~(\ref{eq:unitarity})
(with the inequality saturated), tells us that the imaginary part
of the $J=2$ partial wave of this one-loop diagram equals the
square of the $J=2$ partial wave of the tree diagram. Thus the
factor $N$ is simply the weighted sum over all the particles that
circulate in the loop.

The $J=2$ partial-wave amplitude to one-loop order is
\begin{equation}
a_2 = a_2^{(1)}(1+{\rm Re}\; a_2^{(2)}/a_2^{(1)}+ia_2^{(1)})
\end{equation}
where the superscript in parentheses indicates the order in
$G_NE^2$.  We have used ${\rm Im}\; a_2^{(2)}=|a_2^{(1)}|^2$ in
the above equation, as well as the fact that $a_2^{(1)}$, given by
Eq.~(\ref{eq:eigenvalue}), is real. The unitarity condition $|{\rm
Re}\; a_2| \le 1/2$ implies $|a_2^{(1)}|\le 1/2$ at tree level. We
see that this corresponds to the imaginary part of the one-loop
correction being less than half of the tree-level amplitude. This
is a reasonable criterion for the convergence of the perturbative
expansion.

The real part of the one-loop correction, ${\rm Re}\; a_2^{(2)}$,
is ultraviolet divergent.  This divergence is absorbed into the
coefficients of the $R^2$ and $R^{\mu\nu}R_{\mu\nu}$ terms in the
Lagrangian \cite{Donoghue:dn}.  Thus the coefficients of these
terms, relative to the coefficient of the $R$ term in the
Lagrangian of Eq.~(\ref{eq:lagrangian}), are of order $G_NN$,
rather than simply $G_N$.

\begin{figure}[tb]
\begin{center}
    \mbox{\psfig{file=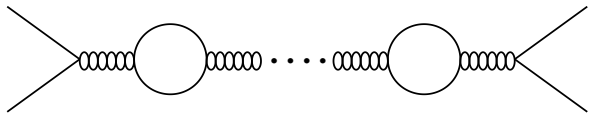,width=8cm} }
\end{center}
    \caption {Multi-loop correction to the scattering of elementary particles
 via $s$-channel graviton exchange in the large-$N$ limit. Each loop contains a sum over all
elementary particles.}
    \label{multi}
\end{figure}

Since $N$ is large, it is interesting to consider the large-$N$
limit of the scattering amplitude, defined as $N\to \infty$ while
keeping $G_N N$ fixed. In this limit, the leading correction to
the scattering amplitude at $n^{\rm th}$ order arises from the
$(n-1)^{\rm th}$ iteration of the one-loop diagram, as shown in
Fig.~\ref{multi} \cite{Tomboulis:jk}. One can sum the geometric
series generated by these leading corrections, to obtain
\begin{equation}
a_2 = \frac{a_2^{(1)}}{1-{\rm Re}\;
a_2^{(2)}/a_2^{(1)}-ia_2^{(1)}}\;. \label{eq:largeN}\end{equation}
This expression for the $J=2$ partial wave satisfies elastic
unitarity, Eq.~(\ref{eq:unitarity}) (with the inequality
saturated), exactly, regardless of the size of $a_2^{(1)}$.  Thus,
at leading order in $1/N$, unitarity is respected at arbitrary
energies, despite the violation of tree unitarity at the energy
given in Eq.~(\ref{bound}).

The amplitude in Eq.~(\ref{eq:largeN}) holds on the positive, real
$s$ axis above the branch cut. More generally, the amplitude in
the complex $s$ plane, in the large-$N$ limit, may be obtained
from Eq.~(\ref{eq:largeN}) by analytic continuation:
\begin{equation}
a_2 =
\frac{a_2^{(1)}}{1+\frac{1}{\pi}a_2^{(1)}\ln\left(-\frac{s}{\mu^2}\right)}\; ,
\end{equation}
where $\mu^2$ is an arbitrary renormalization scale. This
amplitude is unitary, but a pair of complex-conjugate poles on the
physical sheet violate the usual analyticity properties. These
poles yield acausal effects which become appreciable at energies
near $(G_NN)^{-1/2}$  \cite{Tomboulis:jk}.

%

In this paper we have found that tree unitarity is violated in the
effective field theory of quantum gravity coupled to $N_s$
scalars, $N_f$ fermions, and $N_V$ vectors, at an energy squared
of $E_{CM}^2=20(G_NN)^{-1}$, where $N={2\over 3}N_s+N_f+4N_V$.  We
showed that this is related to radiative corrections, proportional
to $G_NNE^2$, due to loops of such particles in the graviton
propagator. New physics must enter before $E_{CM}\approx 6 \times
10^{18}$ GeV in the standard model, and $4.6 \times 10^{18}$ GeV
in the minimal supersymmetric standard model.

\section*{Acknowledgments}

\indent\indent We are grateful for conversations with D.~Chung,
C.~Goebel, J.~Harvey, A.~Manohar, J.~Polchinski, J.~Rosner,
I.~Rothstein, G.~Valencia, M.~Voloshin, L.~Wang, T.~Weiler, and
E.~Weinberg.  We thank the Aspen Center for Physics for
hospitality.  This work was supported in part by the
U.~S.~Department of Energy under contracts Nos.~DE-FG02-95ER40896
and DE-FG02-91ER40677, by the Wisconsin Alumni Research
Foundation, by the National Science Foundation under Grant
No.~PHY99-07949, and by National Natural Science Foundation of
China.


\end{document}